\documentclass [12pt,epsf]{article}
\usepackage{graphics}
\usepackage{graphicx}

\input{epsf}
\begin{document}
\begin{titlepage}
\begin{center}

 \vspace{-0.1in}

{\large \bf Quantum Bound on the Specific Entropy \\ in \\ Strong-Coupled Scalar Field Theory }\\
\vspace{.5in}{\large\em M. Aparicio Alcalde\,\footnotemark[1], G.
Menezes\,\footnotemark[2]
 and N. F. Svaiter\,\,\footnotemark[3]}\\
\vspace{.1in}

Centro Brasileiro de Pesquisas F\'{i}sicas-CBPF\\
Rua Dr. Xavier Sigaud 150,
Rio de Janeiro, RJ,\,22290-180, Brazil\\

\subsection*{\\Abstract}
\end{center}

\baselineskip .10in

We discuss the $(g_{0}\,\varphi^{\,p})_{d}$ self-interacting
scalar field theory, in the strong-coupling regime. We assume the
presence of macroscopic boundaries confining the field in a
hypercube of side $L$. We also consider that the system is in
thermal equilibrium at temperature $\beta^{\,-1}$. For spatially
bounded free fields, the Bekenstein bound states that the specific
entropy satisfies the inequality $\frac{S}{E} < 2 \pi R$, where
$R$ stands for the radius of the smallest sphere that
circumscribes the system. Employing the strong-coupling
perturbative expansion, we obtain the renormalized mean energy $E$
and entropy $S$ for the system up to the order
$(g_{0})^{-\frac{2}{p}}$, presenting an analytical proof that the
specific entropy also satisfies in some situations a quantum
bound. Defining $\varepsilon_d^{(r)}$ as the renormalized
zero-point energy for the free theory per unit length, the
dimensionless quantity $\xi=\frac{\beta}{L}$ and $h_1(d)$ and
$h_2(d)$ as positive analytic functions of $d$, for the case of
high temperature, we get that the specific entropy satisfies
$\frac{S}{E} < 2\pi R\,\frac{h_1(d)}{h_2(d)}\,\xi$. When
considering the low temperature behavior of the specific entropy,
we have $\frac{S}{E} <2\pi R\,
\frac{h_1(d)}{\varepsilon_d^{(r)}}\xi^{\,1-d}$. Therefore the sign
of the renormalized zero-point energy can invalidate this quantum
bound. If the renormalized zero point-energy is a positive
quantity, at intermediate temperatures and in the low temperature
limit, there is a quantum bound.
\vspace{0,3cm}\\
PACS numbers: 03.70+k, 04.62.+v

\footnotetext[1]{e-mail:\,\,aparicio@cbpf.br}
\footnotetext[2]{e-mail:\,\,gsm@cbpf.br}
\footnotetext[3]{e-mail:\,\,nfuxsvai@cbpf.br}

\end{titlepage}
\newpage\baselineskip .18in

\section{Introduction}
\quad \,\, There have been a lot of activities discussing
classical and quantum fields in the presence of macroscopic
boundaries. These subjects raise many interesting questions, since
boundaries introduce a characteristic size in the theory. For
example, in the field-theoretical description of critical
phenomena, the confinement of critical fluctuations of an order
parameter is able to generate long-range forces between the
surfaces of a film. This is known as statistical mechanical
Casimir effect \cite{duplantier} \cite{marcelo} \cite{krech}
\cite{brankov}. These long-range forces in statistical mechanical
systems are characterized by the excess free energy due to the
finite-size contributions to the free energy of the system. It
should be noted that the statistical mechanical Casimir effect is
still waiting for a satisfactory experimental verification. On the
other hand, the electromagnetic Casimir effect \cite{casimir},
where neutral and perfectly conducting parallel plates in vacuum
attract each other, has been tested experimentally with high
accuracy. The introduction of a pair of conducting plates into the
vacuum of the electromagnetic field alters the zero-point
fluctuations of the field and thereby produces an attraction
between the plates \cite{plunien} \cite{mamayev} \cite{bordag}
\cite{milton}. A still open question is how the sign of the
Casimir force depends on the topology, dimensionality of the
spacetime, the shape of bounding geometry or others physical
properties of the system \cite{ambjorn} \cite{caruso}
\cite{robson} \cite{amaral}. We should emphasize that the problem
of the sign of the renormalized zero-point energy of free fields
described by Gaussian functional integrals is crucial for the
subject that we are interested to investigate in this paper.

Another basic question that has been discussed in this scenario,
when quantum fields interact with boundaries, is about the issue
that these systems may be subjected to certain fundamental bounds.
One of these proposed bounds relates the entropy $S$ and the energy
$E$ of the quantum system, respectively, with the size of the
boundaries that confine the fields. This is known as the Bekenstein
bound which is given by $S \leq 2 \pi\,E\, R/\hbar\,c$, where $R$
stands for the radius of the smallest sphere that circumscribes the
system \cite{bek0} \cite{bek1} \cite{bek4} \cite{bek2} \cite{beki5}.
Such bound was originally motivated by considerations of
gravitational entropy, a consistency condition between black hole
thermodynamics and ordinary statistical physics that could guarantee
that the generalized second law of thermodynamics is respected,
which states that the sum of the black-hole entropy and the entropy
of the matter outside the black-hole does not decreases. For
example, in a Schwarschild black-hole in a four-dimensional
spacetime, the Bekenstein entropy, which is proportional to the area
of the spherical symmetric system, exactly saturates the bound. When
gravity is negligible, the bound must be valid for a variety of
systems.

Although analytical proofs of this quantum bound on specific
entropy for free fields has been proposed in the literature, many
authors in the past criticized the bound \cite{count1}
\cite{count2} \cite{count3} \cite{count4} \cite{bouss}. Deutsch
claims that the quantum bound is inapplicable as it stands to
non-gravitating systems, since an absolute value of energy cannot
be observed, and also that for sufficient low temperatures, a
generic system in thermal equilibrium also violates the entropy
bound. Unruh pointed out that for system with zero modes, the
specific entropy cannot satisfy any bound. Many of these
criticisms were answered by Bekenstein and collaborators. The
problem of the low temperature systems was answered in Ref.
\cite{bekn1} and the problem of systems with zero modes was
answered in the Ref. \cite{bekn2}. An strong argument used in one
of these examples is based in the fact that the renormalized
zero-point energy of some free quantum field could be negative.
Some authors claim that, if we take into account the boundaries
responsible for the Casimir energy, it is possible to compensate
their negative energy, yielding a positive total energy which
respects the Bekenstein bound, although this is far from a simple
problem \cite{bousso2}.

In fact this last point that we pointed out, i.e., the objection
raised against the violation of the Bekenstein bound for free
fields, must be analyzed more carefully. The problem now confronting
us is to prove that, although Casimir energy can be negative for
some physical situation, the sum of the energy of the boundary and
the Casimir energy will be positive. In other words, the
contribution from the boundary would make the total energy of the
system always positive for any configuration of the hyperplane when
the Casimir energy is negative. How close the boundaries must be in
such a way that the positive contribution from the rest mass of the
boundary is always smaller than the modulus of the Casimir energy?
Our conclusions from the above arguments is that, for example, in
the case of the electromagnetic field, it is essential to construct
a microscopic model where effects of dispersion and absorbtion must
be taken into consideration. Many authors discussed the problem of
quantization of the electromagnetic field in dispersive and
absorptive linear dielectrics \cite{em1} \cite{em2} \cite{em3}
\cite{em4}. The essential question that confronts us is the
positivity of the total energy of any quantum system defined in a
compact domain in any situation. The validity of the Bekenstein
bound for configurations with negative Casimir energy depends on the
answer for this last question. Our intention in this paper is not to
study such deep and difficult question, introducing physically
realistic boundary conditions, but only to discuss the situation of
idealized mirror boundaries.

We may observe that another quite important situation has not been
discussed systematically in the literature, at least as far we
known. A step that remains to be derived is the validity of the
bound for the case of interacting fields, which are described by
non-Gaussian functional integrals, at least up to some order of the
perturbation theory. Nonlinear interactions can change dramatically
the energy spectrum of the system and this might lead to the
overthrow of the bound \cite{bek3} \cite{int}. The difficulties that
appear in the implementation of this program in the presence of
macroscopic boundaries are well known. In systems where the
translational invariance is lacking it is much harder to compute the
Feynman diagrams then in the unbounded space. Nevertheless a
regularization and renormalization procedure can in principle be
carry out in any order of the perturbative expansion \cite{sy}. See
for example refs. \cite{fosco} \cite{caicedo} \cite{nfs}
\cite{aparicio}, where the perturbative renormalization were
presented in first and second order of the loop expansion in the
$\lambda\varphi^{4}$ self-interacting scalar field theory. We would
like to stress that the renormalization program is implement in a
different way from unbounded or translational invariant systems
because surface divergences appear.

The aim of this paper is to show for a given self-interaction
field theory in which situations the specific entropy satisfies a
quantum bound. For the answer of this important question, there
are two different routes. The first one is to use the
weak-coupling perturbative expansion. However, as we discussed, in
the renormalization procedure appears surface divergences that
forces one to introduce non-local counterterms. It is unclear for
us how this affect the physical relevance of the results that can
be obtained. There are some kind of problems for which the mean
energy and the canonical entropy a system can easily be found for
quantum fields defined in a simple connected bounded region. We
can show that using the strong-coupling expansion one can evaluate
the mean energy and the canonical entropy of the system in a
regime in which quantum fluctuations dominate.

Therefore we study the $(g_{0}\,\varphi^{\,p})_{d}$
self-interacting scalar field theory in the strong-coupling
regime. We assume the presence of macroscopic boundaries that
confine the field in a hypercube of side $L$ and also that the
system is in thermal equilibrium with a reservoir. We present an
analytic proof that, up to the order $(g_{0})^{-\frac{2}{p}}$, the
specific entropy satisfies in some situations a quantum bound.
Defining $\varepsilon_d^{(r)}$ as the renormalized zero-point
energy for the free theory per unit length, $\xi=\frac{\beta}{L}$
and $h_1(d)$ and $h_2(d)$ as positive analytic functions of $d$,
for the case of high temperature, we get that the specific entropy
satisfies the inequality $\frac{S}{E} < 2\pi
R\,\frac{h_1(d)}{h_2(d)}\,\xi$. When considering the low
temperature behavior of the specific entropy, we have $\frac{S}{E}
<2\pi R\, \frac{h_1(d)}{\varepsilon_d^{(r)}}\xi^{\,1-d}$. We are
establishing a bound in the strong-coupled system in the following
cases: in the high temperature limit and if the renormalized zero
point-energy is a positive quantity, at intermediate temperatures
and also in the low temperature limit.

In the weak-coupling perturbative expansion, the information about
the boundaries can be implemented over the free two-point
Schwinger function $G_{0}(m_{0};x-y)$ of the system.  In the
strong-coupling perturbative expansion,  we have to deal with the
problem of how the boundary conditions can be imposed. Let us
briefly discuss the strong-coupling expansion in Euclidean field
theory at zero temperature. The basic idea of the approach is the
following: in a formal representation for the generating
functional of complete Schwinger functions of the theory $Z(V,h)$,
we treat the Gaussian part of the action as a perturbation with
respect to the remaining terms of the functional integral, i.e.,
in the case for the $(g_{0}\,\varphi^{\,p})_{d}$ theory, the local
self-interacting part, in the functional integral. In the
generating functional of complete Schwinger functions, $V$ is the
volume of the Euclidean space where the fields are defined and
$h(x)$ is an external source. We are developing our perturbative
expansion around the independent-value generating functional
$Q_{0}(h)$ \cite{sol1} \cite{meni} \cite{kla2}. In the zero-order
approximation, different points of the Euclidean space are
decoupled since the gradient terms are dropped \cite{kovesi}
\cite{be1} \cite{novo1} \cite{jrk}.

The fundamental problem of the strong-coupling expansion is how to
give meaning to the independent-value generating functional and to
this representation for the Schwinger functional. One attempt is
to replace the Euclidean space by a lattice made by hypercubes. A
naive use of the continuum limit of the lattice regularization,
where one simply makes use of the central limit theorem for the
independent-value generating functional, leads to a Gaussian
theory. A solution to this problem was presented by Klauder a long
time ago \cite{sol1} \cite{kla2} \cite{jrk}. The modification
which allows us to avoid this limitation is a change in the usual
definition of the measure in the functional integral, which
possesses local translational invariance, by another one which is
non-translational invariant.

Let us remark that, in the strong-coupling regime, assuming that the
source is constant, we can perform the perturbative expansion around
a independent-value generating function, up to the order
$(g_{0})^{-\frac{2}{p}}$, and it is possible to split $\ln Z(V,h)$
in two contributions: one that contains only the independent-value
generating function and other that contains the spectral
zeta-function. Therefore, in order to obtain the thermodynamic
quantities, one must proceed in two stages. First, one gives a
operational meaning to the independent-value generating function;
then, one consistently implements the boundary conditions in the
strong-coupling regime. Since we are working in first order of
perturbation theory, to implement boundary conditions, we use the
spectral zeta-function method \cite{hawking} \cite{voros}. For a
complete review about the subject see for example refs. \cite{zeta1}
\cite{zeta2}. Quite recently a very simple application of this
formalism was presented \cite{novo2}, where it was considered an
anharmonic oscillator in thermal equilibrium with a reservoir at
temperature $\beta^{-1}$. Using the strong-coupling expansion, it
was found the mean energy in the regime $\lambda\,\gg\,\omega$, up
to the order $\frac{1}{\sqrt{\lambda}}$, where $\lambda$ and
$\omega$ are the coupling constant and the frequency of the
oscillator respectively.

The organization of the paper is as follows: In section II we
discuss the strong-coupling expansion for the
$(g_{0}\,\varphi^{\,p})_{d}$ theory. In section III we discuss he
free energy and the spectral zeta-function of the system. In
section IV we show that it is possible to obtain in some
situations a quantum bound in the considered model. Finally,
section V contains our conclusions. In the appendix A we present
the Klauder's result, as the formal definition of the
independent-value generating functional derived for scalar fields
in a $d$-dimensional Euclidean space. In the appendix B we prove
that the spectral zeta-function $\zeta_{D}(s)$ evaluated in the
extended complex plane at $s=0$ vanishes. To simplify the
calculations we assume the units to be such that
$\hbar=c=k_{B}=1$.

\section{The strong-coupling perturbative expansion for
scalar $(g_{0}\,\varphi^{\,p})_{d}$ theory}\

Let us consider a neutral scalar field with a
$(g_{0}\,\varphi^{\,p})$ self-interaction, defined in a
$d$-dimensional Minkowski spacetime. The vacuum persistence
functional is the generating functional of all vacuum expectation
value of time-ordered products of the theory. The Euclidean field
theory can be obtained by analytic continuation to imaginary time
allowed by the positive energy condition for the relativistic
field theory. In the Euclidean field theory, we have the Euclidean
counterpart for the vacuum persistence functional, that is, the
generating functional of complete Schwinger functions. In a
$d$-dimensional Euclidean space, the self-interaction contribution
to the action is given by
\begin{equation}
S_{I}(\varphi)= \int d^{\,d}x\,\frac{g_{0}}{p\,!}
\,\varphi^{\,p}(x). \label{3}
\end{equation}

The basic idea of the strong-coupling expansion at zero
temperature is to treat the Gaussian part of the action as a
perturbation with respect to the remaining terms of the action in
the functional integral. Let us assume a compact Euclidean space
with or without a boundary, where the volume of the Euclidean
space is $V$. Let us suppose that there exists an elliptic,
semi-positive, and self-adjoint differential operator $O$ acting
on scalar functions on the Euclidean space. The usual example is
$O=(-\Delta+m_{0}^{2}\,)$, where $\Delta$ is the $d$-dimensional
Laplacian. The kernel $K(m_{0};\, x,y)\equiv\,K(m_{0};\,x-y)$ is
defined by
\begin{equation}
K(m_{0};x-y)=\left(-\Delta+m_{0}^{2}\,\right)\delta^{d}(x-y).
\label{kernel2}
\end{equation}
Using the fact that the functional integral which defines $Z(V,h)$
is invariant with respect to the choice of the quadratic part, let
us consider a modification of the strong-coupling expansion. We
split the quadratic part in the functional integral which is
proportional to the mass squared in two parts; one in the
derivative terms of the action, and the other in the independent
value generating functional. The Schwinger functional can be
defined by a new formal expression for the functional integral
given by
\begin{equation}
Z(V,h)=\exp\left(-\frac{1}{2}\int d^{d}x\,\int
d^{d}y\frac{\delta}{\delta h(x)}K(m_{0},\sigma;x-y)
\frac{\delta}{\delta h(y)}\right)\,Q_{0}(\sigma,h), \label{18}
\end{equation}
where  $Q_{0}(\sigma,h)$, the new independent value functional
integral, is given by
\begin{equation}
Q_{0}(\sigma,h)={\cal{N}}\int [d\varphi]\,\exp \Biggl(\int
d^{d}x\,\biggl(-\frac{1}{2}\,\sigma\, m_{0}^{2}\,\varphi^{2}(x)-
\,\frac{g_{0}}{p\,!}\,\varphi^{p}(x)
+h(x)\varphi(x)\biggr)\Biggr), \label{19}
\end{equation}
and the modified kernel $K(m_{0},\sigma;x-y)$ that appears in Eq.
(\ref{18}), is defined by
\begin{equation}
K(m_{0},\sigma;x-y)=\left(-\Delta+(1-\sigma)m_{0}^{2}\,\right)\delta^{d}(x-y),
\label{benar2}
\end{equation}
where $\sigma$ is a complex parameter defined in the region
$0\leq\,\mbox{Re}\,\,(\sigma)< 1$.

The factor ${\cal{N}}$ is a normalization that can be found using
that $Q_{0}(\sigma,h)|_{\,h=0}=1$. Observe that the non-derivative
terms which are non-Gaussian in the original action do appear in
the functional integral that defines $Q_{0}(\sigma,h)$. At this
point it is convenient to consider $h(x)$ to be complex.
Consequently $h(x)=\mbox{Re}(h)+i\,\mbox{Im}(h)$. In the paper we
are concerned with the case $\mbox{Re}(h)=0$.

Since we are assuming a spatially bounded system in equilibrium
with a thermal reservoir at temperature $\beta^{-1}$, the
strong-coupling expansion can be used to compute the partition
function defined by $Z(\beta,\Omega,h)|_{\,h=0}$, where $h$ is a
external source and we are defining the volume of the $(d-1)$
manifold as $V_{d-1}\equiv\,\Omega$. From the partition function
we define the free energy of the system, given by
$F(\beta,\Omega)=-\frac{1}{\beta}\ln\,Z(\beta,\Omega,h)|_{\,h=0}$.
This quantity can be used to derive the mean energy
$E(\beta,\Omega)$, defined as
\begin{equation}
E (\beta,\Omega) = - \frac{\partial }{\partial\beta}\ln
Z(\beta,\Omega,h)|_{\,h=0}, \label{imp}
\end{equation}
and the canonical entropy $S(\beta,\Omega)$ of the system in
equilibrium with a reservoir with a finite size given by
\begin{equation}
S (\beta,\Omega)= \biggl(1 - \beta \frac{\partial}{\partial\beta}
\biggr)\ln Z(\beta,\Omega,h)|_{h=0}. \label{imp1}
\end{equation}

In the next section we will show that in a particular situation it
is possible, up to the order $(g_{0})^{-\frac{2}{p}}$ to split
$\ln Z(\beta,\Omega,h)$ in two parts: the first one that contains
only the independent-value generating function and the second one
that has the information on the boundary condition and it is given
by derivative of the spectral zeta-function defined in the
extended complex plane in $s=0$ .

\section{The independent-value generating function and
the spectral zeta-function}

We are interested in global quantities. For simplicity we are
assuming that the external source $h(x)$ is constant. In this
situation we call $Z(V,h)$ as a generating function. Since we are
introducing boundaries in the domain where the field is defined,
the spectrum of the operator
$D=\left(-\Delta+(1-\sigma)m_{0}^{2}\,\right)$ has a denumerable
contribution, and an analytic regularization procedure can be used
to control the divergences of the theory. In order to impose
boundary conditions the functional integral must be taken over
functions restricted to the geometric configurations. At zero
temperature, in the leading-order approximation (up to the to the
order $(g_{0})^{-\frac{2}{p}}$) we can write the logarithm of the
generating function as
\begin{equation}
\ln
Z(\beta,\Omega,h)=\frac{1}{Q_{0}(\sigma,h)}
\frac{\partial^{2}}{\partial\,h^{2}}\,Q_{0}(\sigma,
h)
\left(-\frac{\alpha}{2}+\frac{1}{2}
\frac{d}{ds}\,\zeta_{D}(s)|_{\,s=0}\right),
\label{zeta1}
\end{equation}
where $\alpha$ is a infinite constant and $\zeta_{D}(s)$ is the
spectral zeta-function associated with the elliptic operator $D$.

Let us consider now the situation in which the system is finite
along each one of the spatial dimensions, i.e., $x_i \in
\,\,[0,L]$, $i = 1, 2, ..., d - 1 $. For the Euclidean time we
assume periodic boundary conditions (Kubo-Martin-Schwinger KMS
\cite{kubo} \cite{martin} conditions) and for the Euclidean
spatial dimensions we assume Dirichlet boundary conditions. We
call this latter situation "hard" boundaries. See for example the
Ref. \cite{ss}. For different kinds of confining boundaries see
\cite{caruso1} \cite{fluct}. The choice of the hard boundary
provides an easy solution to the eigenvalue problem, so that
explicit and complete calculation using the spectral-zeta function
can be performed without difficulty.

It follows that the operator $D$ has the spectrum given by
$\lambda_{\,n_1,\,...\,,\,n_d}\,$ where
\begin{equation}
\lambda_{\,n_1,\,...\,,\,n_d}\,=\,\biggl[\biggl(\frac{
n_1\pi}{L}\biggr)^{2}+...+\biggl(\frac{ n_{d-1}\pi
}{L}\biggr)^{2}+\biggl(\frac{ 2 n_d\,\pi
}{\beta}\biggr)^{2}+(1-\sigma)m_{0}^{2}\,\biggr], \label{e5}
\end{equation}
$n_{1}, n_{2},...\,,n_{d-1}$ are natural numbers different from
zero, since we are choosing Dirichlet boundary conditions and $n_d$
are integer numbers. The spectral zeta-function associated with the
operator $D$ in this situation reads
\begin{equation}
\zeta_{D}(s)=\,\sum_{n_1,...,\,n_d}^{\infty\,\,\,\,\,_{'}}
\lambda_{\,n_1,...,\,n_d}^{-s}\,, \label{e6}
\end{equation}
where $s$ is a complex parameter, and the prime sign means that the
term $n_{1}=0,n_{2}=0,..,n_{d-1}=0$ must be excluded. The series
above converges for Re$\,s> \,\frac{d}{2}$ and its analytic
continuation defines a meromorphic function of $s$, analytic at
$s=0$. Since we should have to introduce an arbitrary parameter
$\mu$ with dimension of a mass to implement the analytic procedure
with dimensionless quantities, we have scaling properties.

Using $n$ as a general index instead of $n_1,...,\,n_d$, the
scaling properties follows from the fact that
\begin{equation}
\zeta_{\,\mu D}(s)= \sum_{n}^{\infty}
(\,\mu^{-2}\,\lambda_{\,n})^{-s}=\mu^{2s}\sum_{n}^{\infty}
\lambda_{\,n}^{-s}=\mu^{2s}\zeta_{D}(s)\,. \label{bound7}
\end{equation}
Therefore we have
\begin{equation}
\frac{1}{2} \frac{d}{ds}\,\zeta_{\,\mu D}(s)|_{\,s=0}= \frac{1}{2}
\frac{d}{ds}\,\zeta_{D}(s)|_{\,s=0}
+\frac{1}{2}\ln\mu^2\,\zeta_{D} (s)|_{\,s=0}. \label{ref3}
\end{equation}

Before continuing, we would like to discuss two points. The first
one is the fact that for different boundary condition, as, for
example, Neumann boundary conditions in all the hyperplanes or
periodic boundary conditions in all the spatial directions, the
presence of the zero-mode can make the calculations more involved.
Studying a two-dimensional spacetime $T\times S^{1}$ and also
$T\times S^{3}$, Dowker \cite{referee21} discussed how the spatial
zero mode contributes to the free energy. The zero-mode problem
was also discussed by Dowker and Kirsten \cite{referee22}.
Elizalde and Tort \cite{referee23} discussed the contribution to
the thermal energy coming from the spatial zero mode in a system
defined in a manifold with non-trivial topology; being more
specific these authors studied a massive scalar field in
$S^{1}\times S^{3}$ geometry. Kirsten and Elizalde discussed the
Casimir energy of a massive scalar field in a general
$(2+1)$\,-\,dimensional toroidal spacetime. For the massless case
they excluded the $n_{1}=n_{2}=0$ mode \cite{kireli}. They
discussed also in an ultrastatic $d$-dimensional spacetime how the
ambiguity of the Casimir energy is related to the
$B_{\frac{d}{2}}$.  The entropy bound related to various fields in
the $R\times\,S^{3}$ geometry was also discussed by many authors.
See for example the  Refs. \cite{referee24} \cite{referee25}. It
is important to remark that this zero mode problem does not appear
in the calculations that we are presenting, since we are choosing
Dirichlet boundary conditions in all hyperplanes, excluding the
possibility of the spatial zero mode. The generalized Bekenstein
bound for systems with the spatial zero mode must be taken in
consideration to extend the results of the paper.

The second point is that it is possible to show that there is no
scaling in the situation that we are interested in. The spectral
zeta-function is related to the heat-kernel or diffusion operator
via a Mellin transform. The trace of the diffusion operator is the
integral of the diagonal part of the heat-kernel over the
manifold. It is possible to perform an asymptotic expansion for
the heat-kernel and this asymptotic expansion shows that the
spectral zeta-function is a meromorphic function of the complex
variable $s$  possessing simple poles where the residues of the
poles depends on the $B_{n}$ coefficients which depends on the
Seeley-DeWitt coefficients, the second fundamental form on the
boundary and the induced geometry on the boundary. See for example
the Ref. \cite{moss} \cite{bor1}. It is possible to show that the
polar structure of the analytic extension of the spectral zeta
function in a compact manifold with boundary is given by
\begin{equation}
\zeta_{D}(s)=\frac{1}{(4\pi)^{\frac{d}{2}}}\frac{1}{\Gamma(s)}
\Bigg[\sum_{n=0}^{\infty}\frac{B_{n}}{n-\frac{d}{2}+s}+g_{2}(s)\Biggr],
\label{ref31}
\end{equation}
for $n$ integer or odd-half integer, where $g_{2}(s)$ is an
analytic function in ${\bf C}$. As was stressed by Blau et al
\cite{blau}, in a four dimensional flat spacetime with massless
particles and thin boundaries the geometric coefficient $B_{2}$
vanishes. The spectral zeta function has poles at $s=1$, $s=2$,
and so on. There would be a pole at $s=0$, but is cancelled out by
the pole in the gamma function. Therefore
$\zeta_{D}(s)|_{s=0}=B_{\frac{d}{2}}$. For the case of hypercube
with Dirichlet boundary conditions we prove in the appendix B that
the spectral zeta-function in $s=0$ is zero, consequently
$B_{\frac{d}{2}}=0$ and there is no scaling in the theory. We
shall come back to this point in the conclusions.

Let us study in Eq. (\ref{zeta1}) the contribution arising from
the spectral zeta-function which takes into account the geometric
constraints upon the scalar field. Using the spectrum of the $D$
operator given by Eq. (\ref{e5}) and the definition of the
spectral zeta-function given by Eq. (\ref{e6}), we get that the
derivative of the spectral zeta-function in $s=0$ yields
\begin{equation}
\frac{d}{ds}\,\zeta_{D}(s)|_{\,s=0} = -\,\sum_{\vec{n}_{d-1} =
1}^{\infty}\sum_{ n_d = -\infty}^{\infty}\Biggl(\ln
\biggl(\biggl(\frac{\pi\,\beta\, q }{L}\biggr)^{2} + (2\pi n_d
)^{2}\biggr)+ \ln \Biggl(1 + \frac{a^2\beta^2}{4n_d^2L^2 +
q^2\beta^2}\Biggr)\Biggr), \label{deriva}
\end{equation}
where $\vec{n}_{d-1} = (n_1, n_2, ..., n_{d-1})$, $q^2 = n_{1}^{2}
+  n_{2}^{2} + ... + n_{d-1}^{2}$ and
$a^2=\biggl(\frac{(1-\sigma)m_{0}^{2}L^2}{\pi^2}\biggr)$. Note
that in Eq. (\ref{deriva}) we are using that
$\zeta_{D}(s)|_{\,s=0}=0$. Using the following identity
\cite{kapusta}
\begin{equation}
\ln\Biggl(\biggl(\frac{\pi\,\beta\,q }{L}\biggr)^{2}+(2\pi n_d)^2
\Biggr) = \int_{1}^{(\frac{\pi\,\beta\,q
}{L})^{2}}\,\frac{d\theta^2}{\theta^2 + (2\pi n_d)^2} + \ln\Bigl(1
+ (2\pi n_d)^2\Bigr)\,,  \label{k}
\end{equation}
we can see that the first term in the right hand side of Eq.
(\ref{deriva}) gives a divergent contribution. To proceed we use
another useful identity given by
\begin{equation}
\sum_{n_d = -\infty}^{\infty}\,\frac{1}{\theta^2+(2\pi n_{d})^{2}}
= \frac{1}{2\theta}\biggl(1 + \frac{2}{e^{\theta} - 1}\biggr).
\label{ident}
\end{equation}
Using both identities given by Eq. (\ref{k}) and Eq.
(\ref{ident}), it is possible to express the double summation that
appears in Eq. (\ref{deriva}) by a single summation given by
\begin{equation}
\sum_{\vec{n}_{d-1} = 1}^{\infty}\sum_{ n_d = -\infty}^{\infty}\ln
\Biggl(\biggl(\frac{\pi\,\beta\, q }{L}\biggr)^{2} + (2\pi
n_d)^{2}\Biggr) = 2\sum_{\vec{n}_{d-1} =
1}^{\infty}\int_{1}^{\bigl(\frac{\pi\,\beta\,q
}{L}\bigr)}\,d\theta \biggl(\frac{1}{2} + \frac{1}{e^{\theta} - 1}
\biggr)+ \alpha_{1}, \label{falta}
\end{equation}
where $\alpha_1=\sum_{\vec{n}_{d-1} = 1}^{\infty}\sum_{ n_d =
-\infty}^{\infty}\ln\Bigl(1 + (2\pi n_d)^2\Bigr)$. Carrying out
the $\theta$ integration, we finally arrive that Eq. (\ref{falta})
can be written as
\begin{equation}
\sum_{\vec{n}_{d-1} = 1}^{\infty}\sum_{ n_d = -\infty}^{\infty}\ln
\Biggl(\biggl(\frac{\pi\,\beta\, q }{L}\biggr)^{2} + (2\pi
n_d)^{2}\Biggr) = 2\sum_{\vec{n}_{d-1} =
1}^{\infty}\Biggl(\frac{\pi\,\beta\,q }{2L} + \ln\biggl(1 -
e^{-\frac{\pi\,\beta\,q }{L}}\biggr)\Biggr)+\alpha_2\,, \label{k2}
\end{equation}
where $\alpha_2=\alpha_1-\sum_{\vec{n}_{d-1} =
1}^{\infty}\Bigl(1+2\ln(1-e^{-1})\Bigr)$. Since this divergent
contribution $\alpha_2$ is $\beta$-independent we will see that it
can be eliminated using the third law of thermodynamics. The first
term on the right side of Eq. (\ref{k2}) is a divergent
contribution, corresponding to the zero-point energy term.
Using the following mathematical result \cite{grads} \cite{prud}
given by
\begin{equation}
\prod_{n=-\infty}^{\infty}\Biggl(1 + \frac{a^2}{n^2 + b^2}\Biggr)
= \frac{\sinh^2(\pi\sqrt{a^2 + b^2})}{\sinh^2(\pi\,b)}\,,
\label{id}
\end{equation}
we can write the last term of Eq. (\ref{deriva}) in a more
manageable way. Using the Eq. (\ref{k2}) and Eq. (\ref{id}), the
derivative of the spectral zeta-function in $s=0$ can be rewritten
as
\begin{equation}
\frac{d}{ds}\,\zeta_{D}(s)|_{\,s=0} = - 2\sum_{\vec{n}_{d-1} =
1}^{\infty}\left[\ln
\left(\frac{\sinh\Bigl(\frac{\pi\beta}{2L}\sqrt{q^2 +a^2}\,\Bigr)}
{\sinh\Bigl(\frac{\pi\beta q}{2L}\Bigr)}\right)+ \ln\biggl(1 -
e^{-\frac{\pi\,\beta\,q}{L}}\biggr)+\frac{\pi\,\beta\,q}{2L}\right]-\alpha_2\,.
\label{zeta2}
\end{equation}
It is possible to show (see appendix A) that, in the finite
temperature case, the independent-value generating function
$Q_{0}(\sigma,h)$ satisfies $Q_{0}(\sigma,h)|_{_{h=\sigma=0}}=1$,
and
\begin{equation}
\frac{\partial^{2}}{\partial\,h^{2}}\,Q_{0}(\sigma,
h)|_{\,h=\sigma=0}=
\frac{\Gamma(\frac{2}{p})}{2p\,g_{0}^{\frac{2}{p}}(p\,!)^{\frac{p}{2}}}.
\label{45e}
\end{equation}
In the next section we show  that it is possible to obtain a
quantum bound in the spatially bounded system defined by a
self-interacting scalar field in the strong-coupling regime in
high temperatures. As we will see, for the cases of intermediate
or low temperatures, the sign of the renormalized zero-point
energy is crucial for the validity of a quantum bound for the
specific entropy.

\section{The specific entropy for strongly coupled $(g_{0}\,\varphi^{\,p})_{d}$ theory}

In this section we compute the specific entropy $\frac{S}{E}$ of
the system. For simplicity, let us define $\ln
Z(\beta,\Omega,h)|_{\,h=0}=\ln Z(\beta,\Omega)$. From Eq.
(\ref{imp}) and Eq. (\ref{imp1}), and using for simplicity that
the mean energy $E(\beta,\Omega)= E$ and the entropy
$S(\beta,\Omega)= S$, the specific entropy is given by
\begin{equation}
\frac{S}{E} =\beta - {\ln
Z(\beta,\Omega)}\biggl(\frac{d}{d\beta}\ln
Z(\beta,\Omega)\biggr)^{-1}\,. \label{imp2}
\end{equation}
Substituting Eq. (\ref{zeta2}) and Eq. (\ref{45e}) in Eq.
(\ref{zeta1}) we have that $\ln Z(\beta,\Omega)$ is given by
\begin{equation}
\ln Z(\beta,\Omega)= -
\frac{\Gamma(\frac{2}{p})}{2p\,(p\,!)^{\frac{p}{2}}g_{0}^{\frac{2}{p}}}
\left(\frac{\alpha\,'}{2}+I_2(\beta)\right)\,, \label{451}
\end{equation}
where $\alpha\,'=\alpha+\alpha_2$ and the quantity $I_{2}(\beta)$
is given by
\begin{equation}
I_{2}(\beta) = \sum_{\vec{n}_{d-1} = 1}^{\infty}\left[\ln
\left(\frac{\sinh\Bigl(\frac{\pi\beta}{2L}\sqrt{q^2 +a^2}\,\Bigr)}
{\sinh\Bigl(\frac{\pi\beta q}{2L}\Bigr)}\right)+ \ln\biggl(1 -
e^{-\frac{\pi\,\beta\,q}{L}}\biggr)+\frac{\pi\,\beta\,q}{2L}\right]\,.
\label{zetau}
\end{equation}
Defining $C_1$ and $C_2=\,-\frac{2C_{1}}{\alpha\,'}$ that depend
only of $p$ and $g_{0}$ and do not depend on $\beta$ as
\begin{equation}
C_1=
-\frac{\alpha\,'\,\Gamma(\frac{2}{p})}{4p\,(p\,!)^
{\frac{p}{2}}g_{0}^{\frac{2}{p}}}\,,
\label{referee6}
\end{equation}
the quantity $\ln Z(\beta,\Omega)$ can be written in a general
form as
\begin{equation}
\ln Z(\beta,\Omega)= C_1 - C_2\, I_2(\beta). \label{452}
\end{equation}
It is worth to mention that the quantity $C_1$ corresponds to a
divergent expression, $C_2$ is finite and the summation term in
the right-hand side of Eq. (\ref{zeta2}) is proportional to the
zero-point energy. In order to renormalize $\ln Z(\beta,\Omega)$
we first can use the third law of thermodynamics. The derivative
of $\ln Z(\beta,\Omega)$ with respect of $\beta$ yields
\begin{equation}
\frac{d}{d\beta}\ln Z(\beta,\Omega) = -C_2
\,\frac{d}{d\beta}I_2(\beta), \label{derlogz1}
\end{equation}
where the derivative of $I_{2}(\beta)$ with respect to $\beta$ is
given by
\begin{equation}
\frac{d}{d\beta}I_2(\beta) = \frac{\pi}{2L}\sum_{\vec{n}_{d-1} =
1}^{\infty}\left(\Biggl(\sqrt{q^2+a^2}\,\coth\biggl(\frac{\pi\beta}{2L}
\sqrt{q^2 +a^2}\,\biggr) -
q\,\coth\biggl(\frac{\pi\beta\,q}{2L}\biggr)\Biggr) +
\frac{2\,q}{e^{\frac{\pi\,\beta\, q}{L}} - 1} +q\right)\,.
\label{derlogz3}
\end{equation}
Using the definition of the mean energy given by Eq. (\ref{imp}),
Eq. (\ref{derlogz1}) and Eq. (\ref{derlogz3}) we have that the
unrenormalized mean energy is given by
\begin{equation}
E = \frac{\pi\,C_{2}}{2L}\sum_{\vec{n}_{d-1} =
1}^{\infty}\left(\Biggl(\sqrt{q^2+a^2}\coth\biggl(\frac{\pi\beta}{2L}
\sqrt{q^2 +a^2}\,\biggr) -
q\,\coth\biggl(\frac{\pi\beta\,q}{2L}\biggr)\Biggr) +
\frac{2\,q}{e^{\frac{\pi\,\beta\, q}{L}} - 1} +q\right)\,.
\label{referee1}
\end{equation}
For the case $a=0$, i.e., the massless case, using the Eq.
(\ref{referee6}) we get that the unrenormalized mean energy
$E(\beta,\Omega)$ becomes
\begin{equation}
E(\beta,\Omega)|_{a=0}
=\frac{\pi\,\Gamma(\frac{2}{p})}{4p\,(p\,!)^{\frac{p}{2}}
\,g_{0}^{\frac{2}{p}}L}\,\sum_{\vec{n}_{d-1}
= 1}^{\infty}\left(\frac{2\,q}{e^{\frac{\pi\,\beta\, q}{L}} - 1}
+q\right)\,. \label{referee2}
\end{equation}
The formula above has the simple interpretation of being phase
space sums over the mean energy of each mode, where the zero-point
energy is included. Note that the divergence that appear in the
mean energy given by the Eq. (\ref{referee2}) is coming from the
zero-point energy, which is given by
\begin{equation}
E_0= \frac{\pi}{2L}\sum_{\vec{n}_{d-1} = 1}^{\infty} (n_{1}^{2} +
n_{2}^{2} + ... + n_{d-1}^{2})^{\frac{1}{\,2}}\,, \label{referee3}
\end{equation}
and its sign is given by the ratio between the first and the
second terms of the right-hand side of Eq. (\ref{referee2}), for a
negative zero-point energy. The sign of the renormalized mean
energy will be discussed in details later in this section.

Let us briefly discuss how to find the renormalized zero-point
energy. An analytic regularization gives the renormalized
zero-point energy.  Using the definition of the Epstein-zeta
function given by
\begin{equation}
A(a_{1},a_{2},...,a_{k}; 2s)=\sum_{\vec{n}_{k} = -\infty}^{\infty}
(a_{1}n_{1}^{2} +a_{2} n_{2}^{2} + ... + a_{k}n_{k}^{2})^{-s}\,,
\label{referee4}
\end{equation}
we can find the analytic extension of the Epstein-zeta function in
the complex plane, in particular at $s=-\frac{1}{2}$, to define the
Casimir energy. The structure of the divergences of the analytic
extension of the Epstein-zeta function is well known in the
literature \cite{ambjorn} \cite{anal1} \cite{anal2} \cite{anal3}.
The renormalized zero-point energy $E_{0}^{\,r}$ is defined as the
finite part of a meromorphic function that possesses simple poles.
Details of this calculations can be found in refs. \cite{caruso}
\cite{sinal}. As stressed in the Ref. \cite {blau} there is an
ambiguity in the renormalization procedure to find the Casimir
energy.

It is possible to prove that, in a regularization of an ill
defined quantity, if the introduction of a exponential cut-off
yields a analytical function with a pole in the origin, the
analytic extension of the generalized zeta-function, or the
zeta-function method, is equivalent to the application of a
cut-off with the subtraction of the singular part at the origin
\cite{nami1} \cite{nami2}. Once we accept the advantage of the
zeta-function method over the cut-off method with the subtraction
of the polar terms, due to the fact that it is an analytical
extension method, we face a problem: the non-trivial scaling
behavior of the vacuum energy. This follows from the fact that in
an analytic extension method it is necessary to introduce a mass
parameter $\mu$, i.e., a normalization scale to keep the
Epstein-zeta function dimensionless for all values of $s$. If we
consider a change in the normalization scale
$\mu\rightarrow\,\mu'$, it is possible to show that in a
$d$-dimensional spacetime, the ambiguity of the renormalized
zero-point energy is related to the $B_{\frac{d}{2}}$ coefficient
by the expression:
\begin{equation}
E_{0}^{\,r}(\mu')=E_{0}^{\,r}(\mu)-\mu\,\frac{B_{\frac{d}{2}}}
{(4\pi)^{d/2}}\,\ln\Biggl(\frac{\mu'}{\mu}\Biggr).
\label{referee4}
\end{equation}
Although, in general situations, there is an ambiguity in the
renormalization procedure, in our case
$\zeta_{D}(s)|_{\,s=0}=B_{\frac{d}{2}}=0$. Therefore there is no
scaling in the theory and consequently the renormalized zero-point
energy does not depend on the renormalized scale $\mu$.

Note that although in the expression for the renormalized mean
energy, up to the order $(g_{0})^{-\frac{2}{p}}$, the coupling
constant appears, we are interested only in the ratio
$\frac{S}{E}$, and the dependence of the coupling constant
disappears. It is important to stress that this happens only
because in the strong-coupling expansion, up to the order
$(g_{0})^{-\frac{2}{p}}$, we are able to split $\ln Z$ into two
contributions: the first one proportional to the spectral-zeta
function and the second one that has a contribution from the
independent-value generating function $Q_{0}(\sigma,h)$.

After this discussions we are able to present the entropy of the
system. Substituting Eq. (\ref{452}) and Eq. (\ref{derlogz1}) in
the definition of the entropy given by Eq. (\ref{imp1}), we have
that the entropy of the system can be written as
\begin{equation}
S = C_1 - \beta C_2\left(\frac{I_2(\beta)}{\beta } -
\frac{d}{d\beta}I_2(\beta)\right)\,.
 \label{entropia1}
\end{equation}
A system with a unique ground state corresponds to a state of
vanishing entropy at zero temperature. For systems with degenerate
ground states this property is also valid if the entropy is a
extensive quantity. Since at zero temperature the system goes to a
non-degenerate ground state, the entropy must go to zero. The
expression of the entropy given by Eq. (\ref{entropia1}) must
satisfy the third law of thermodynamics, i.e., the entropy of a
system has a limiting property that
$\lim_{\beta\rightarrow\infty}S=0$. To proceed, lets analyze the
limit given by
\begin{equation}
\lim_{\beta \rightarrow \infty} \frac{I_2(\beta)}{\beta} =
\lim_{\beta \rightarrow \infty} \frac{d}{d\beta}I_2(\beta) =
\frac{\pi a^2}{2L}\sum_{\vec{n}_{d-1} =
1}^{\infty}\frac{1}{\sqrt{q^2 +a^2}\, +
q}+\frac{\pi}{2L}\sum_{\vec{n}_{d-1} = 1}^{\infty}q\,.
\label{lims}
\end{equation}
Substituting Eq. (\ref{lims}) in Eq. (\ref{entropia1}), and using
the third law of thermodynamics, we get
\begin{equation}
\lim_{\beta \rightarrow \infty}S = C_1 =\,0\,.
\end{equation}
Therefore the first step to find a finite result for $\ln
Z(\beta,\Omega)$ was achieved, since we were able to renormalize
$C_{1}$ to zero using the third law of thermodynamics. After this
step we have
\begin{equation}
\ln Z(\beta,\Omega) =-C_2\,I_2(\beta)\,. \label{log}
\end{equation}
Substituting Eq. (\ref{log}) in Eq. (\ref{imp2}) we can see that
for the case $a=0$, i.e., the massless case, the quotient
$\frac{S}{E}$ yields
\begin{equation}
\frac{S}{E} = 2\pi R\, T_d(\xi)\,, \label{specific0}
\end{equation}
where we are defining the dimensionless variable $\xi$ given by
$\xi=\beta/L$. Since the field is confined in a hypercube, the
radius of the smallest $(d-1)$-dimensional sphere that
circumscribes this system should be given by $R =
\frac{1}{2}\sqrt{(d-1)}\,L$. The function $T_d(\xi)$ defined in
Eq. (\ref{specific0}) is given by
\begin{equation}
T_d(\xi) = \frac{1}{\pi\sqrt{d-1}}\,\frac{\xi\,P_d(\xi)+R_d(\xi)}
{\varepsilon_d^{(r)}+P_d(\xi)}\,, \label{specific}
\end{equation}
where $\varepsilon_d^{(r)}=LE_0^{(r)}$ and the positive functions
$P_d(\xi)$ and $R_d(\xi)$ are defined respectively by
\begin{equation}
P_d(\xi) = \sum_{\vec{n}_{d-1} =
1}^{\infty}\pi\,q\left(e^{\pi\,\xi\,q} - 1\right)^{-1} \label{p2}
\end{equation}
and
\begin{equation}
R_d(\xi) = -\sum_{\vec{n}_{d-1} = 1}^{\infty}\ln\left(1 -
e^{-\pi\,\xi\, q}\right)\,. \label{q2}
\end{equation}
Now let us study the function $T_d(\xi)$ given by Eq.
(\ref{specific0}). The quantum bound holds whenever $T_d(\xi)\leq
1$ for all values of $\xi$. From the definition of the function
$T_d(\xi)$, given by Eq. (\ref{specific}), we have that $T_d(\xi)$
has a divergent value only if the renormalized zero-point energy
is negative. For the point $\xi=\xi_0$ which satisfies
$\varepsilon_d^{(r)}+P_d(\xi_0)=0$, the quantum bound is
invalidated.

Numerical calculations can help us understand the quantum bound.
In the Fig. (1) we present the plot of the function $T_d(\xi)$ in
the case of $d=3$ over the interval $0<\xi<2$. Since the
renormalized zero-point energy is positive \cite{sinal}, the
function $T_d(\xi)$ is also positive for all values of $\xi$.
There is a maximum for some value of $\xi$ that we are calling
$\xi_{max}$, which is near one. For this case there is a quantum
bound. In Fig.(2) we present the function $T_d(\xi)$ in the case
of $d=4$ over the interval $0<\xi<2$. Since in this case the
renormalized zero-point energy is negative, we have that for some
value of $\xi=\xi_0$, the function $T_d(\xi)$ diverges. There
exists a critical value $\xi_c$ where, for $\xi\,>\,\xi_c$, the
specific entropy is unbounded above.

Let us analyze two cases. The first one is when the renormalized
zero-point energy is positive (see Fig. 1) and a maximum value for
$T_d(\xi)$ appears. The second case, with a negative renormalized
zero-point energy, invalidate the quantum bound. For even
dimensional spacetime, the renormalized zero-point energy is
always negative. For the odd dimensional case, it is known that
for $d\leq 29$ this quantity is positive and for $d>29$ it changes
the sign \cite{caruso}.

For the cases of positive renormalized zero-point energy, an
equation for the maximum value of $T_d(\xi)$ can be found. The
equation for the maximum is given by
$R_d(\xi_{max})=\varepsilon_d^{(r)}\,\xi_{max}$. Substituting this
$\xi_{max}$ in Eq. (\ref{specific}) we can find that
$T_d(\xi_{max})=\frac{\xi_{max}}{\pi\sqrt{d-1}}$. Using the same
procedure in Eq. (\ref{specific0}) we get
$\frac{S}{E}=\beta_{max}$, where $\beta_{max}=L\,\xi_{max}$.
Therefore we can conclude that for odd space-time dimensions
$d\leq 29$ there exists a maximum value for the function
$T_d(\xi)$.

We can see that the maximum value of $T_d(\xi)$ depends on the
renormalized zero-point energy, where for the case $d=3$ is less
than one. To prove that, for odd $d\leq\,29$, we have that
$T_d(\xi)$ satisfies the inequality $T_d(\xi)<1$, let us define an
auxiliary function $R\,'_d(\xi)$ that satisfies $R_d(\xi) <
R\,'_d(\xi)$. This function is given by
\begin{equation}
R\,'_d(\xi) = - \int_{\Omega_R}d\Omega_{d-1}
\int_{0}^{\infty}\,dr\,r^{d-2} \ln\left(1 - e^{-\pi\,\xi\,
r}\right)\,, \label{q1}
\end{equation}
where the angular domain of integration $\Omega_R$ correspond to
the region where $r_i>0$. Performing this integral \cite{prud} we
have that
\begin{equation}
R\,'_d(\xi) = S_{d-1}\,\Gamma(d-1)\,\zeta(d)\biggl(\frac{1}{\pi
\xi}\biggr)^{d-1}\,. \label{q4}
\end{equation}
where the angular term is
$S_{d-1}=\frac{(\sqrt{\pi})^{d-1}}{2^{d-2}\Gamma(\frac{d-1}{2})}$.
Using the Eq. (\ref{q4}) in the equation for the maximum, i.e.,
$R_d(\xi_{max})=\varepsilon_d^{(r)}\,\xi_{max}$, we can find that
$\xi_{max}<\xi'_{max}$, where
\begin{equation}
\xi'_{max}=\left(\frac{2}{(2\sqrt{\pi})^{d-1}}\frac{\Gamma(d-1)\,\zeta(d)}
{\Gamma(\frac{d-1}{2})\,\varepsilon_d^{(r)}}\right)^{\frac{1}{d}}\,,
\end{equation}
and we have that
$T_d(\xi_{max})<\frac{\xi'_{max}}{\pi\sqrt{d-1}}$. In the table 1
we present the maximum values for $d=3$ until $d=29$ for odd
$d$'s.
\begin{center}
\begin{tabular}{|c||c|c|c|c|c|c|}
\hline {\bf d} & {\bf 3} & {\bf 5} & {\bf 7} & {\bf 9} & {\bf 11} & {\bf 13} \\
 \hline \hline
$\varepsilon_d^{(r)}$ & $4.1\times 10^{-2}$ & $6.2\times 10^{-3}$
& $1.1\times 10^{-3}$ & $2.2\times 10^{-4}$ & $4.4\times 10^{-5}$
& $9.4\times 10^{-6}$ \\
 \hline
 $T_d(\xi_{max})<$ & 0.3763 & 0.2645 & 0.2303 & 0.2130 &
0.2025 & 0.1953 \\
 \hline
\end{tabular}
\end{center}
\begin{center}
\begin{tabular}{|c||c|c|c|c|c|c|}
\hline {\bf d} &  {\bf 15} & {\bf 17} & {\bf 19} & {\bf 21} & {\bf 23} & {\bf 25} \\
 \hline \hline
$\varepsilon_d^{(r)}$ & $2.0\times 10^{-6}$ & $4.5\times 10^{-7}$
& $1.0\times 10^{-8}$ & $2.2\times 10^{-8}$ & $5.0\times 10^{-9}$
& $1.1\times 10^{-9}$ \\
 \hline
 $T_d(\xi_{max})<$ & 0.1901 & 0.1861 & 0.1829 & 0.1804 & 0.1784 & 0.1769 \\
 \hline
\end{tabular}
\end{center}
\begin{center}
\begin{tabular}{|c||c|c|c|}
\hline {\bf d} & {\bf 27} & {\bf 29} & {\bf 31}\\
 \hline \hline
$\varepsilon_d^{(r)}$ &  $2.3\times 10^{-10}$ &
$3.0\times 10^{-11}$ & $-1.1\times 10^{-11}$\\
 \hline
 $T_d(\xi_{max})<$ & 0.1761 & 0.1781 & no maximum\\
 \hline
\end{tabular}
\end{center}
\begin{figure}[h]
\begin{center}
\includegraphics[height=8cm]{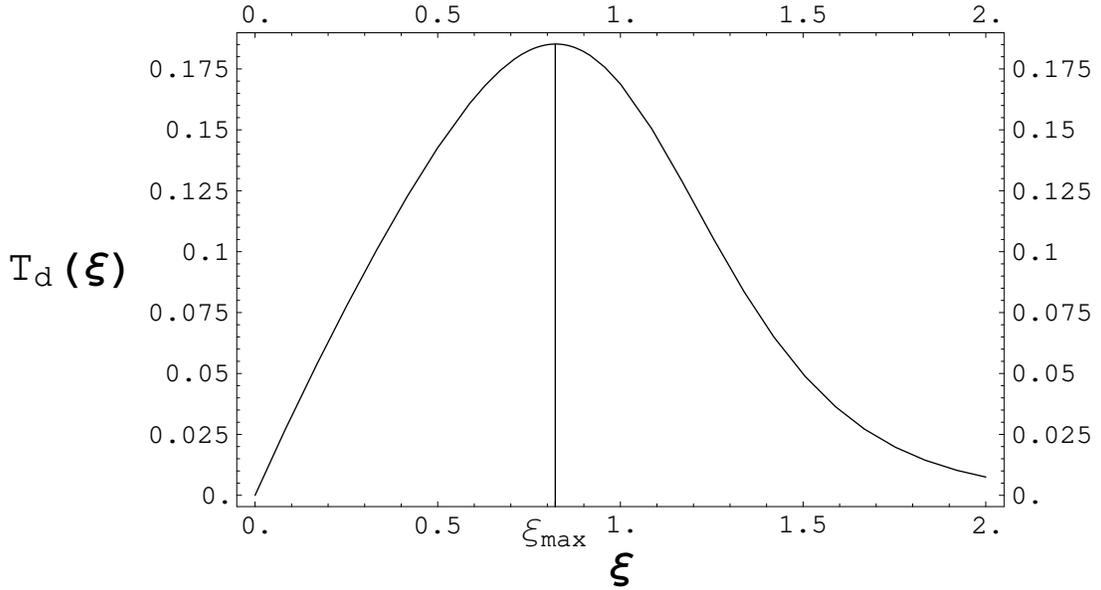}
\caption{$T_d(\xi)$ as a function of $\xi$ for the case of
positive renormalized zero-point energy for $d=3$.}
 \end{center}
\end{figure}

\begin{figure}[h]
\begin{center}
\includegraphics[height=8cm]{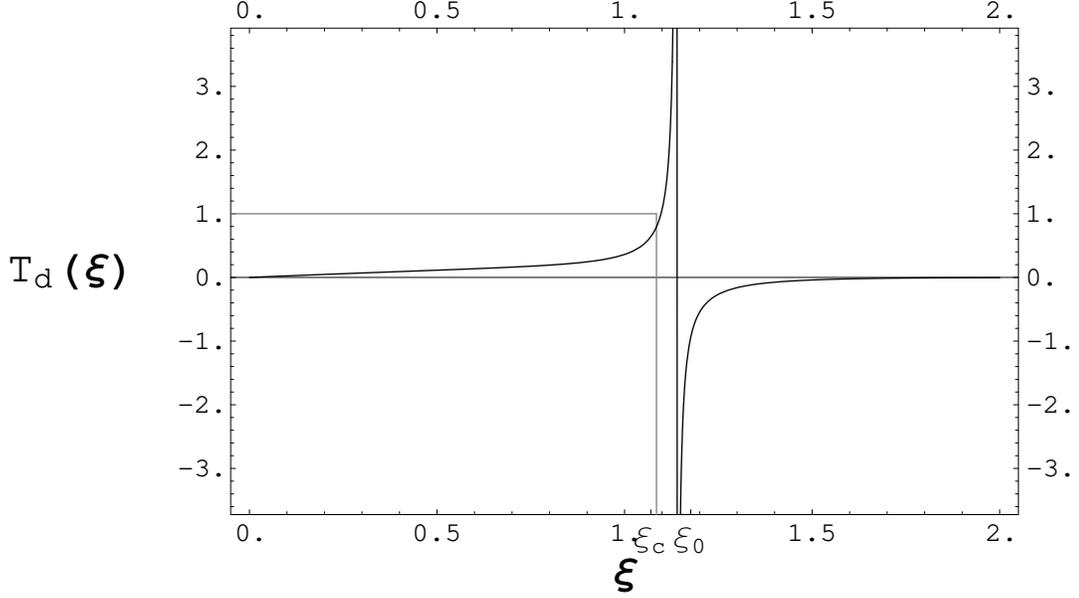}
\caption{$T_d(\xi)$ as a function of $\xi$ for the case of
negative renormalized zero-point energy for $d=4$.}
\end{center}
\end{figure}
Until now we studied the quantum bound for general dimensions
based on the summations given by Eq. (\ref{p2}) and Eq.
(\ref{q2}). Nevertheless we can find an upper bound function
$T\,'_d(\xi)$ of the function $T_d(\xi)$ which is more manageable.
For this purpose, in a similar way as we have defined the function
$R\,'_d(\xi)$, let us define also the auxiliary functions
$P\,'_d(\xi)$ and $P\,''_d(\xi)$, that satisfy
\begin{equation}
P_d(\xi) < P\,'_d(\xi)\,,
\end{equation}
and
\begin{equation}
P_d(\xi) > P\,''_d(\xi)\,,
\end{equation}
so that the specific entropy satisfies the following inequality
\begin{equation}
\frac{S}{E} < 2\pi R\, T\,'_d(\xi)\,, \label{specific1}
\end{equation}
where
\begin{equation}
T\,'_d(\xi) =
\frac{1}{\pi\sqrt{d-1}}\,\frac{\xi\,P\,'_d(\xi)+R\,'_d(\xi)}
{\varepsilon_d^{(r)}+P\,''_d(\xi)}\,. \label{specific11}
\end{equation}
Without loss of generality we can choose as the auxiliary
functions $P\,'_d(\xi)$ and $P\,''_d(\xi)$ by the integrals
\begin{equation}
P\,'_d(\xi) = \pi\,\int_{\Omega_R}d\Omega_{d-1}
\int_{0}^{\infty}\,dr\,r^{d-1}\left(e^{\pi\,\xi\, r} -
1\right)^{-1} \label{p11}
\end{equation}
and
\begin{equation}
P\,''_d(\xi) = \pi\,\int_{\Omega_R}d\Omega_{d-1}
\int_{1}^{\infty}\,dr\,r^{d-1}\left(e^{\pi\,\xi\, r} -
1\right)^{-1}\,. \label{p1}
\end{equation}
Performing these integrals \cite{prud}, we obtain that
$P\,'_d(\xi)$ and $P\,''_d(\xi)$ are given by
\begin{equation}
P\,'_d(\xi) = \pi S_{d-1} \Gamma(d)\zeta(d) \biggl(\frac{1}{\pi
\xi}\biggr)^{d} \label{p13}
\end{equation}
and
\begin{equation}
P\,''_d(\xi) = \pi S_{d-1} \Bigl(\Gamma(d)\zeta(d) -
f(d)\Bigr)\biggl(\frac{1}{\pi \xi}\biggr)^{d}\,, \label{p3}
\end{equation}
where the series $f(d)$ is given by
\begin{equation}
f(d) = \sum_{l = 0}^{\infty}\frac{B_{l}}{(d + l - 1)l!}.
\end{equation}
To obtain an upper bound for the specific entropy in a generic
Euclidean $d$-dimensional spacetime we have only to substitute Eq.
(\ref{q4}), Eq. (\ref{p13}) and Eq. (\ref{p3}) into Eq.
(\ref{specific11}). We have that
\begin{equation}
T\,'_d(\xi)= \frac{h_1(d)}
{\varepsilon_d^{(r)}\,\xi^{d-1}+h_2(d)\,\xi^{-1}}\,,
\end{equation}
where
\begin{equation}
h_{1}(d) =
\frac{S_{d-1}}{\pi^{d}\sqrt{d-1}}\zeta(d)\biggl(\Gamma(d)+\Gamma(d-1)\biggr)\,,
\label{h1}
\end{equation}
and
\begin{equation}
h_{2}(d) = \frac{S_{d-1}}{\pi^{d-1}}
\biggl(\Gamma(d)\,\zeta(d)-f(d)\biggr)\,. \label{h2}
\end{equation}

It is interesting to study the behavior of the specific entropy
for low and high temperatures. For the case of high temperatures,
we get
\begin{equation}
\frac{S}{E} < 2\pi R\,\frac{h_1(d)}{h_2(d)}\,\xi\, .
\end{equation}
At high temperatures the dimension in the imaginary direction
shrinks to zero and the system behaves like a classical system in
$(d-1)$ dimensions where quantum fluctuations are absent. This
behavior of the specific entropy increasing with $\beta$ in the
high-temperature limit was obtained by Deutsch in Ref.
\cite{count3}. Bekenstein using the condition $\beta \ll R$ (high
temperature limit) also obtained the same behavior in Ref.
\cite{beki5}. Since the thermal energy can compensate the negative
renormalized zero-point energy, the quantum bound holds.

When considering the low temperature behavior of the specific
entropy, we can see that the problem of the sign of the
renormalized zero-point energy can invalidate the quantum bound.
In this limit we have
\begin{equation}
\frac{S}{E} <2\pi R\,
\frac{h_1(d)}{\varepsilon_d^{(r)}}\xi^{1-d}\,.
\end{equation}
Although some authors claim that the energy of the boundaries of
such systems can compensate the negative renormalized-zero point
energy yielding a net positive energy, for us this is still an
open question that deserves further investigation. Note that
although our results are based in a quite particular choice of the
shape of the macroscopic boundaries that confine the field in the
volume $\Omega$, it is tempting to think that, using some results
from spectral geometry \cite{luis}, the quantum bound could be
generalized to more general geometries.

\section{Conclusions and perspectives}

\quad In this paper we studied self-interacting scalar fields in
the strong-coupling regime in equilibrium with a thermal bath,
also in the presence of macroscopic boundaries. In the
strong-coupling perturbative expansion we may split the problem of
defining the generating functional in two parts: how to define
precisely the independent-value generating functional and how to
go beyond the independent-value approximation, taking into account
the perturbation part. The presence of the spectral zeta-function
allow us to introduce the boundary conditions in the problem.
Using the Klauder representation for the independent-value
generating functional, and up to the order
$(g_{0})^{-\frac{2}{p}}$, we show that it is possible to obtain a
quantum bound in the system defined by a self-interacting scalar
field in the strong-coupling regime. We established a bound on
information storage capacity of the strong-coupled system in a
framework independent of gravitational physics.

We have shown that, in the strong-coupling regime, at low and
intermediate temperatures $(\beta\approx L)$, the quantum bound
depends on the sign of the renormalized zero-point energy given by
$E_0^{(r)}$. For even spacetime dimensions $d$ and also for odd
values satisfying the inequality $d\,>\,29$, $E_0^{(r)}$ is a
negative quantity. Therefore the quantum bound is invalidated. For
odd values of $d$, satisfying the inequality $d\,\leq\,29$,
$E_0^{(r)}$ is a positive quantity. In  this situation the
specific entropy satisfies a quantum bound. Defining
$\varepsilon_d^{(r)}$ as the renormalized zero-point energy for
the free theory per unit length, we get the following functional
dependencies. For low temperatures we get  $\frac{S}{E} <2\pi R\,
\frac{h_1(d)}{\varepsilon_d^{(r)}\xi^{\,d-1}}$, where $R$ is the
radius of the smallest sphere circumscribing the system. For the
case of high temperature, we get that the specific entropy always
satisfies a quantum bound, given by $\frac{S}{E} < 2\pi
R\,\frac{h_1(d)}{h_2(d)}\,\xi$.

Before finishing, we would like to discuss whether the additive
normalization energy is acceptable for an entropy-energy bound, in
respect that at principle we expect that such rate must be
independent of any additive normalization. Let us discuss briefly
the normalization condition, since we have the ambiguity in the
finite part of the renormalized zero-point energy. As we
discussed, a merit of the zeta function method over the cut-off
method is the fact that it is an analytic extension method, where
we should introduce a mass parameter to keep the generalized zeta
function of the problem a dimensionless quantity for all values of
$s$. We would like to stress that this is a general feature of any
method which is based in the principle of analytic continuation.
The conclusions of the argument become obvious. We have to
consider the effect of a change in the normalization scale.
Therefore we have scaling properties. These scaling properties are
given in Eq. (\ref{ref3}). We proved that the spectral
zeta-function in $s=0$ is zero and from Eq. (\ref{ref31}) we have
that $\zeta(s)|_{s=0}=B_{\frac{d}{2}}$. The conclusion is that in
the compact region that we are confining the field
$B_{\frac{d}{2}} = 0$. Now we are able to discuss the ambiguity in
the finite part of the renormalized quantities which belongs to
the free theories, i.e., the zero-point energy. As we discussed,
if we consider a change in the renormalization scale $\mu$, it can
be shown that the Casimir energy defined in a $d$-dimensional
spacetime associated with the scale $\mu$ and the scale $\mu'$ are
related by the expression given by Eq. (\ref{referee4}). Since
$B_{\frac{d}{2}} = 0$, we can claim that our results are presented
in a normalization independent way.

We would like to stress that without a proof that the spectral
zeta-function in $s=0$ is zero, the correction coming from the
boundaries and interaction at principle are not present in a
normalized independent way, since we have an ambiguity in the
zero-point energy. This ambiguity was discussed by many authors in
distinct situations. For example, for the case of massive scalar
field theory in a classical background field, Bordag, Mohideen and
Mostepanenko \cite{bordag} claim that, in the limit of infinite
mass, quantum fluctuations must vanish, so the renormalized energy
must vanish as well. This was also discussed in refs.  \cite
{bord} \cite{vass}. In a four dimensional spacetime, in the
massless case, it was shown that there is no general normalization
condition, if the $B_{2}$ coefficient is non-zero.

Although in the framework of analytic extension procedures there is
no clear resolution at the present to the ambiguity for the
zero-point energy if $B_{\frac{d}{2}}\neq\,0$, the global energy of
the force between the hyperplanes has an unambiguous value. We
should note that using an alternative procedure, the ambiguity of
the additive normalization can be fixed in the following way: in the
regularization procedure, using a exponential cut-off, with
subtraction of configurations, if the boundaries go to infinity, the
renormalized zero-point energy must be zero. Important arguments
supporting this procedure are based in the demonstration that the
expected value of the energy-momentum tensor in the vacuum state
(since it is a state belonging to discrete spectrum, normalized to
unit and Poincar\'e invariant) should be zero to ensure that the
correct commutation relations of the Lie algebra are satisfied by
the generators of the Poincar\'e group \cite{taka}.  The conclusion
is that, even after obtaining a finite result for the vacuum energy,
we still have to use a physical argument to fix the value of the
energy for some configuration. Therefore our results are acceptable
for an entropy-energy bound, in respect that such rate must be
uniquely defined after raising the ambiguity in the finite part of
the renormalized zero-point energy.

There are some continuations for this paper. Still using the scalar
field, the Bekenstein bound should be investigated assuming more
general boundary conditions over the macroscopic boundaries that
confine the field. Another interesting point is to investigate the
Bekenstein bound in different quantum field models still using the
strong-coupling expansion. For an alternative method to investigate
the strong-coupling regime in quantum field theory, see for example
Ref.\cite{efimov}. The generalization of the Bekenstein bound in a
theory with vector and spinor field is under investigation by the
authors \cite{proximo}.

\begin{appendix}
\makeatletter \@addtoreset{equation}{section} \makeatother
\renewcommand{\theequation}{\thesection.\arabic{equation}}

\section{Appendix:\,The Klauder representation for the independent value generating
functional}

To give meaning to the independent value generating functional
$Q_{0}(\sigma,h)$, we are using the Klauder's result as the formal
definition of the independent-value generating functional derived
for scalar fields in a $d$-dimensional Euclidean space. It is
possible to show that the independent-value generating function
can be written as
\begin{equation}
Q_{0}(\sigma,h)=\exp\Biggl(-\frac{1}{2V}\int\,d^{d}x
\int_{-\infty}^{\infty}\frac{du}{|u|}\left(1-\cos(hu)\right)\exp
\left(-\frac{1}{2}\,\sigma\,m_{0}^{2}\,u^{2}-
\frac{g_{0}}{p\,!}\,u^{p}\right)\Biggl). \label{353}
\end{equation}
There is no need to go into details of this derivation (see Ref.
\cite{jrk}). We would like to point out that in Klauder's
derivation for the free independent-value model a result was
obtained which is well defined for all functions which are square
integrable in $R^{\,n}$ i.e., $h(x)\,\, \in\,L^{2}(R^{\,n})$.
Since we are assuming that $h=cte$, we have to normalize our
expressions. In order to study $Q_{0}(\sigma,h)$ let us define
$E(m_{0},\sigma,g_{0},h)$ given by
\begin{equation}
E(m_{0},\sigma,g_{0},h)=
\int_{-\infty}^{\infty}\frac{du}{|u|}\left(1-\cos(hu)\right)\exp\left(
-\frac{1}{2}\,\sigma\,m_{0}^{2}\,u^{2}-
\frac{g_{0}}{p\,!}\,u^{p}\right). \label{aa}
\end{equation}
Using a series representation for $\cos x$ and using the fact that
the series obtained $(\sum_{k=1}^{\infty}c_{k}\,f_{k}(u)$) not
only converges on the interval $[0,\infty)$, but also converges
uniformly there, the series can be integrated term by term. It is
not difficult to show that
\begin{equation}
E(m_{0},\sigma,g_{0},h)=
2\sum_{k=1}^{\infty}\frac{(-1)^{k}}{(2k)!}h^{2k}
\int_{0}^{\infty}du\,u^{2k-1}\exp
\left(-\frac{1}{2}\,\sigma\,m_{0}^{2}\,u^{2}-
\frac{g_{0}}{p\,!}\,u^{p}\right). \label{38}
\end{equation}
Now let us use the fact that the $\sigma$ parameter can be choose
in such a way that the calculations becomes tractable. Let us
choose $\sigma=0$. Therefore we have
\begin{equation}
E(m_{0},\sigma,g_{0},h)|_{\,\sigma=0}=
2\sum_{k=1}^{\infty}\frac{(-1)^{k}}{2k!}h^{2k}
\int_{0}^{\infty}du\,u^{2k-1}\exp(-\frac{g_{0}}{p\,!}\,u^{p}).
\label{39}
\end{equation}
Let us use the following integral representation for the Gamma
function \cite{grads}
\begin{equation}
\int_{0}^{\infty}\,dx\,x^{\nu-1}\exp(-\mu\, x^{p})=\frac{1}{p}
\,\mu^{-\frac{\nu}{p}}\,\Gamma\left(\frac{\nu}{p}\right),\,\,\,\,\,
\mbox{Re}(\mu)>0\,\,\,\,\,\,\,\mbox{Re}(\nu)>0\,\,\,\,p>0.
\label{40}
\end{equation}
At this point it is clear that the $(g_{0}\,\varphi^{p})$ theory,
for even $p>4$, can easily handle applying our method. Using the
result given by Eq. (\ref{40}) in Eq. (\ref{39})  we have
\begin{equation}
E(m_{0},\sigma,g_{0},h)|_{\,\sigma=0}=\sum_{k=1}^{\infty}
g(p,k)\frac{h^{2k}}{g_{0}^\frac{2 k}{p}}, \label{41}
\end{equation}
where the coefficients $g(p,k)$ are given by
\begin{equation}
g(p,k)=\frac{2}{p}\frac{(-1)^{k}}{(2k)!}(p\,!)^{\frac{2
k}{p}}\Gamma(\frac{2 k}{p}). \label{42}
\end{equation}
Substituting the Eq. (\ref{41}) and Eq. (\ref{42}) in Eq.
(\ref{353}) we obtain that the independent-value generating
function $Q_{0}(\sigma,h)|_{\,\sigma=0}$ can be written as
\begin{equation}
Q_{0}(\sigma,h)|_{\,\sigma=0}=\exp\Biggl[-\frac{1}{2
\Omega\beta}\,\int_{0}^{\beta}\,d\tau \int
d^{d-1}x\,\sum_{k=1}^{\infty}\,g(p,k)\frac{h^{2k}}{g_{0}^\frac{2
k}{p}}\Biggl]. \label{43}
\end{equation}
It is easy to calculate the second derivative for the
independent-value generating function with respect to $h$. Note
that $Q_{0}(\sigma,h)|_{_{\,h=\sigma=0}}=1$. Thus we have
\begin{equation}
\frac{\partial^{2}}{\partial\,h^{2}}\,Q_{0}(\sigma,h)|_{\,\sigma=0}=
\left(-\frac{1}{2}
\sum_{k=1}^{\infty}\,g(p,k)(2k)(2k-1)\frac{h^{2k-2}}{g_{0}^\frac{2
k}{p}}\right) \exp\left(-\frac{1}{2}
\sum_{k=1}^{\infty}\,g(p,k)\frac{h^{2k}}{g_{0}^\frac{2
k}{p}}\right) +G(g_{0}, p, h), \label{44}
\end{equation}
where $G(g_{0},p,h)$ is given by
\begin{equation}
G(g_{0},p,h)=\left(\sum_{k,\,q
=1}^{\infty}\,g(p,k,q)\frac{h^{2k+2q-2}}{g_{0}^\frac{2(k+q)}{p}}\right)
\exp\left(-\frac{1}{2}
\sum_{k=1}^{\infty}\,g(p,k)\frac{h^{2k}}{g_{0}^\frac{2
k}{p}}\right), \label{iii}
\end{equation}
and $g(k,q)=k\,q\,g(k)g(q)$. We are interested in the case $h=0$,
therefore the double series does not contribute to the Eq.
(\ref{44}), since $lim_{h\rightarrow 0} G(h)=0$. Using the fact
that we are interested in the case $h=0$, we have the simple
result that in the Eq. (\ref{44}) only the term $k=1$ contributes.
We get
\begin{equation}
\frac{\partial^{2}}{\partial\,h^{2}}\,Q_{0}(\sigma,
h)|_{\,h=\sigma=0}=\frac{\Gamma(\frac{2}{p})}{2p\,g_{0}^{\frac{2}{p}}(p\,!)^{\frac{p}{2}}}.
\label{45}
\end{equation}

\section{Appendix: \,Proof that the value of the spectral zeta-function in
the origin vanishes, i.e., $\zeta_{D}(s)|_{s=0}=0$}

As we discussed before, to take into account the scaling
properties we should have to introduce an arbitrary parameter
$\mu$ with dimension of a mass to define all the dimensionless
physical quantities and in particular make the change
\begin{equation}
\frac{1}{2} \frac{d}{ds}\,\zeta_{D}(s)|_{\,s=0}\rightarrow
\frac{1}{2} \frac{d}{ds}\,\zeta_{D}(s)|_{\,s=0}
-\frac{1}{2}\ln\bigg(\frac{1}{4\pi\mu^{2}}\bigg)\zeta_{D}
(s)|_{\,s=0}.
\end{equation}
In this appendix we have a proof that the spectral zeta-function
in $s=0$ is zero, consequently there is no scaling in the theory.
The Epstein zeta-function is defined by
\begin{equation}
Z_p\,(a_1,...,a_p\,;2s)=\sum_{n_1,...,\,n_p\,=-\infty}^{~~\infty\,\,,}
\left((a_1\,n_1)^2+...+(a_p\,n_p)^2\right)^{-s}\,,
\label{epstein1}
\end{equation}
where the prime indicates that the term for which all $n_i=0$ is
to be omitted. This summation is convergent only for $2s>p$.
Nevertheless, we can find an integral representation which gives
an analytic continuation for the Epstein zeta-function except for
a pole at $p=2s$ \cite{ambjorn}. This representation is given by
\begin{eqnarray}
&&(\pi\,\eta)^{-s}\,\Gamma(s)\,Z_p\,(a_1,...,a_p\,;2s)=\nonumber\\
&&-\frac{1}{s}+\frac{2}{p-2s}+\eta^{-s}\int_{\eta}^{\infty}dx\,x^{s-1}
\left(\vartheta(0,...,0;a_1^2\,x,...,a_p^2\,x)-1\right)\nonumber\\
&&+\eta^{(2s-p)/2}\int_{1/\eta}^{\infty}dx\,x^{(p-2s)/2-1}
\left(\vartheta(0,...,0;x/a_1^2,...,x/a_p^2)-1\right)\,,
\label{epstein2}
\end{eqnarray}
where $\eta^{\,p/2}$ is the product of the $p\,'s$ parameters
$a_{i}$ given by $\eta^{\,p/2}=a_1...a_p$, and the generalized
Jacobi function $\vartheta(z_1,...,z_p;x_1,...,x_p)$, is defined
by
\begin{equation} \vartheta(z_1,...,z_p\,;x_1,...,x_p)=\prod_{i=1}^{p}\vartheta(z_i;x_i)\,,
\end{equation}
with $\vartheta(z;x)$ being the Jacobi function, i.e.,
\begin{equation}
\vartheta(z;x)=\sum_{n=-\infty}^{\infty}e^{\pi(2nz-n^2x)}\,.
\end{equation}
Using this integral expression for the Epstein zeta-function,
given by Eq. (\ref{epstein2}), we can find that
\begin{equation}
Z_p(a_1,...,a_p\,;2s)|_{s=0}=-1\,, \label{epstein3}
\end{equation}
for any $p\geq 1$. To proceed, let us define the function
$Z_p^{(q)}(a_1,...,a_p\,;2s)$, given by
\begin{equation}
Z_p^{(q)}(a_1,...,a_p\,;2s)=\sum_{n_1,...,\,n_q=1}^{\infty}\,\,\sum_{n_{q+1},...,n_p\,=-\infty}^{\infty}
\left((a_1\,n_1)^2+...+(a_p\,n_p)^2\right)^{-s}\,.
\label{q-epstein1}
\end{equation}
Using the result given in Eq. (\ref{epstein3}) we can show that,
after performing the analytic continuation of the function
$Z_p^{(q)}(a_1,...,a_p\,;2s)$, the following property holds
\begin{equation}
Z_p^{(q)}(a_1,...,a_p;2s)|_{s=0}=0\,,
\label{q-epstein2}
\end{equation}
where this result is valid only for $0<q<p$. We can prove this
property by induction. First, let us verify that for $q=1$ the
above property hold. Therefore, assuming that is valid for $q$, we
have only to show that is true for $q+1$. For $q=1$ we have that
\begin{equation}
Z_p(a_1,...,a_p\,;2s)|_{\,s=0}=Z_p(a_2,...,a_p\,;2s)|_{\,s=0}
+2Z_p^{(1)}(a_1,...,a_p\,;2s)|_{\,s=0}\,. \label{pr1}
\end{equation}
Since $p>1$ we can use the property given by Eq. (\ref{epstein3}),
for the two first terms of Eq. (\ref{pr1}) and verify that
$Z_p^{(1)}(a_1,...,a_p\,;2s) |_{s=0}=0$. The next step in the
proof by induction is to assume the validity of this property for
some $q$, i.e., $Z_p^{(q)}(a_1,...,a_p\,;2s)|_{\,s=0}=0$, with $p$
being arbitrary, but satisfying the condition $0<q<p$, then we
must verify the validity of this property for $q+1$, i.e.,
$Z_{p\,'}^{(q+1)}(a_1,...,a_{p\,'}\,;2s)|_{\,s=0}=0$ with $p\,'$
also being arbitrary but satisfying the condition $0<q+1<p\,'$.
From the following property
\begin{equation}
Z_{p\,'}^{(q)}(a_1,...,a_{p\,'}\,;2s)|_{s=0}=Z_{p\,'-1}^{(q)}
(a_1,...,a_q,a_{q+2},...,a_{p\,'};2s)|_{s=0}
+2Z_{p\,'}^{(q+1)}(a_1,...,a_{p\,'};2s)|_{\,s=0}\,, \label{pr2}
\end{equation}
since $0<q<p\,'-1$ and using the assumption of the validity of
this property for arbitrary $q$, given by Eq. (\ref{q-epstein2}),
we can see that the two first terms in Eq. (\ref{pr2}) vanish.
Therefore we finally proved that
$Z_{p\,'}^{(q+1)}(a_1,...,a_{p\,'}\,;2s)|_{\,s=0}=0$. We are
interested in a particular case of this property, given by
\begin{equation}
Z_p^{(p-1)}(a_1,...,a_p;2s)|_{s=0}=\left(\sum_{n_1,...,\,n_{p-1}=1}^{\infty}
\,\,\sum_{n_p=-\infty}^{\infty}
\left((a_1\,n_1)^2+...+(a_p\,n_p)^2\right)^{-s}\right)\Bigg|_{\,s=0}=0\,.
\label{p-1-epstein}
\end{equation}
\end{appendix}

\section{Acknowlegements}

We would like to thanks M. I. Caicedo and J. Stephany for many
helpful discussions. We are grateful to J. D. Bekenstein for
useful comments that improved the presentation of the paper. N. F.
Svaiter would like to acknowledge the hospitality of the
Departamento de Fisica da Universidade Simon Bolivar, where part
of this paper was carried out. This paper was partially supported
by Conselho Nacional de Desenvolvimento Cientifico e
Tecnol{\'o}gico do Brazil (CNPq).


\begin{thebibliography}{99}

\bibitem{duplantier} A. Ajdari, B. Duplantier, D. Hone, L. Peliti
and J. Prost, J. Phys. II France {\bf 2}, 487 (1992).
\bibitem{marcelo} M. L. Lyra, M. Kardar and N. F. Svaiter, Phys. Rev. {\bf E47}, 3456 (1993).
\bibitem{krech} M. Krech, {\em{''The Casimir Effect in Critical
Systems''}}, World Scientific, Singapure (1994).
\bibitem{brankov} J. G. Brankov, D. M. Danchev and M. S. Tonchev.
{\em{'' Theory of Critical Phenomena in Finite Size Systems''}},
World Scientific, Singapure (2000).
\bibitem{casimir} H. B.  G. Casimir,
Proc. Kon. Ned. Akad. Wekf. {\bf 51}, 793 (1948).
\bibitem{plunien} G. Plunien, B. M\"uller and W. Greiner, Phys. Rep. {\bf 134},
87 (1986).
\bibitem{mamayev} A. A. Grib, S. G. Mamayev and V. M.
Mostepanenko, {\em{"Vacuum Quantum Effects in Strong Fields"}},
Friedman Laboratory Publishing. St. Petesburg (1994).
\bibitem{bordag} M. Bordag, U. Mohideen and V. M. Mostepanenko, Phys. Rep.
{\bf 353}, 1 (2001).
\bibitem{milton} K. A. Milton, {\em{"The Casimir Effect:  Physical
Manifestation of Zero-Point Energy"}}, World Scientific (2001).
\bibitem{ambjorn} J. Ambjorn and S. Wolfram, Ann. Phys. {\bf
147}, 1 (1983).
\bibitem{caruso} F. Caruso, N. P. Neto,  B. F. Svaiter and N. F. Svaiter,
 Phys. Rev. {\bf D43}, 1300 (1991).
\bibitem{robson}
R. D. M. De Paola, R. B. Rodrigues and N. F. Svaiter, Mod. Phys.
Lett. {\bf A34}, 2353 (1999).
\bibitem{amaral} L. E. Oxman. N. F. Svaiter and R. L. P. G.
Amaral, Phys. Rev. {\bf D72}, 125007 (2005).
\bibitem{bek0}J. D. Bekenstein, Phys. Rev. {\bf D7}, 2333 (1973).
\bibitem{bek1} J. D. Bekenstein, Phys. Rev. {\bf D23}, 287 (1981).
\bibitem{bek4} J. D. Bekenstein, Phys. Rev. {\bf D30}, 1669 (1984).
\bibitem{bek2} M. Schiffer and J. D. Bekenstein, Phys. Rev. {\bf D39}, 1109 (1989).
\bibitem{beki5} J. D. Bekenstein, Phys. Rev. {\bf D49}, 1912 (1994).
\bibitem{count1} D. N. Page, Phys. Rev {\bf D26}, 947 (1982).
\bibitem{count2} D. Unwin, Phys. Rev. {\bf D26}, 944 (1982).
\bibitem{count3} D. Deutch, Phys. Rev. Lett. {\bf 48}, 286 (1982).
\bibitem{count4} W. G. Unruh, Phys. Rev. {\bf D42}, 3596 (1990).
\bibitem{bouss} R. Bousso, J. High Energy Phys. {\bf 04}, 035 (2004).
\bibitem{bekn1}J. D. Bekenstein, Foundations of Physics 35, 1805 (2005).
\bibitem{bekn2} M. Schiffer and J. D. Bekenstein, Physical Review {\bf D42},
3598 (1990).
\bibitem{bousso2} R. Bousso, {\em{Bound States and the Bekenstein
Bound}} ArXiv hep-th/0310148 (2003).
\bibitem{em1} B. Huttner and S. M. Barnett, Phys. Rev. {\bf A46}, 4306 (1992).
\bibitem{em2} T. Gruner and D.-G. Welsch, Phys. Rev. {\bf A51}, 3246 (1995).
\bibitem{em3} T. Gruner and D.-G. Welsch, Phys. Rev. {\bf A53}, 1818 (1996).
\bibitem{em4} R. Matloob,  Phys. Rev. {\bf A60}, 50 (1999).
\bibitem{bek3} J. D. Bekenstein and E. I. Guendelman, Phys. Rev. {\bf D35}, 716 (1987).
\bibitem{int} J. D. Bekenstein and M. Schiffer, Int. J. Mod. Phys.
{\bf C1}, 355 (1990).
\bibitem{sy} K. Symanzik, Nucl. Phys. {\bf B190}, 1 (1981).
\bibitem{fosco}  C. D. Fosco and N. F. Svaiter, J. Math. Phys. {\bf 42}, 5185, (2001).
\bibitem{caicedo} M. I. Caicedo and N. F. Svaiter, J. Math.
Phys. {\bf 45}, 179 (2004).
\bibitem{nfs} N. F. Svaiter, J. Math.
Phys. {\bf 45}, 4524 (2004).
\bibitem{aparicio} M. Aparicio Alcalde, G. F. Hidalgo and N. F. Svaiter, J. Math.
Phys. {\bf 47}, 052303 (2006).
\bibitem{sol1} J. R. Klauder, Acta Phys. Aust. {\bf 41}, 237 (1975).
\bibitem{meni} R. Menikoff and D. H. Sharp, J. Math. Phys. {\bf 19}, 135 (1978).
\bibitem{kla2} J. R. Klauder, Ann. Phys. {\bf 117}, 19 (1979).
\bibitem{kovesi} S. Kovesi-Domokos, Il Nuovo Cim. {\bf 33A}, 769 (1976).
\bibitem{be1} C. M. Bender,
F. Cooper, G. S. Guralnik and D. H. Sharp, Phys. Rev. {\bf D19},
1865 (1979).
\bibitem{novo1} N. F. Svaiter, Physica {\bf A345}, 517 (2005).
\bibitem{jrk} J. R. Klauder, {\em{''Beyond Conventional Quantization''}}, Cambridge University
Press, Cambridge (2000).
\bibitem{hawking} S. W. Hawking, Comm. Math. Phys. {\bf 55}, 133 (1977).
\bibitem{voros} A. Voros, Comm. Math. Phys. {\bf 110}, 439 (1987).
\bibitem{zeta1} E. Elizalde, S. D. Odintsov, A. Romeo, A. A.
Bytsenko and S. Zerbini, {\em{''Zeta Regularization Techniques and
Applications''}}, World Scientific, Singapure (1994).
\bibitem{zeta2} K. Kirsten, {\em{''Spectral Functions in Mathematics and Physics''}},
Chapman and Hall/CRC, Florida (2002).
\bibitem{novo2} N. F. Svaiter, Physica {\bf
A386}, 111 (2006).
\bibitem{kubo} R. Kubo, J. Phys. Soc. Jap. {\bf 12}, 570 (1957).
\bibitem{martin} P. C. Martin and J. Schwinger, Phys. Rev. {\bf
115}, 1342 (1959).
\bibitem{ss} N. F. Svaiter and B. F. Svaiter, Jour. Phys. {\bf A25}, 979
(1992).
\bibitem{caruso1} F. Caruso, R. De Paola and N. F. Svaiter, Int. Jour.
Mod. Phys. {\bf A14}, 2077 (1999).
\bibitem{fluct} L. H. Ford and N. F. Svaiter,  Phys. Rev. {\bf 58}, 065007-1 (1998).
\bibitem{referee21} J. S. Dowker, hep-th/0203026
\bibitem{referee22} J. S. Dowker and K. Kirsten, Anal. and Geom.
{\bf 7}, 641 (1999).
\bibitem{referee23} E. Elizalde and A. C. Tort, Phys. Rev. {\bf
D66}, 045033 (2002).
\bibitem{kireli} K. Kirsten and E. Elizalde, Phys. Lett {\bf
B365}, 72 (1996).
\bibitem{referee24} I. Brevik, K. A. Milton and S. D. Odinstov,
Ann. Phys. {\bf 302}, 120 (2002).
\bibitem{referee25} I. Brevik, K. A. Milton and S. D. Odinstov,
hep-th/0210286.
\bibitem{moss} J. G. Moss, Class. Quant. Grav {\bf 6}, 759 (1989).
\bibitem{bor1} M. Bordag, E. Elizalde and K. Kirsten, J. Math.
Phys. {\bf 37}, 895 (1996).
\bibitem{blau} S. K. Blau, M. Visser and A. Wipf, Nucl. Phys. {\bf
B310}, 163 (1988).
\bibitem{kapusta} J. I. Kapusta, {\em{``Finite-temperature
Field Theory''}}, Cambridge University Press (1989).
\bibitem{grads}I. S. Gradshteyn and I. M. Ryzhik, {\em{''Tables of Integrals, Series
and Products''}}, Academic Press Inc., New York (1980).
\bibitem{prud} A. P. Prudnikov, Yu. A. Brychkov, O. I. Marichev, {\em{''Integrals and Series''}},
Vol. 1 and 2, Gordon and Breach Science Publishers (1986).
\bibitem{anal1} L.H. Ford, Phys. Rev. {\bf D21}, 933 (1980).
\bibitem{anal2} K. Kirsten, J. Math. Phys. {\bf 32}, 3008 (1991).
\bibitem{anal3} L. H. Ford and N. F. Svaiter, Phys. Rev. {\bf
D51}, 6981 (1995).
\bibitem{sinal} J. R. Ruggiero, A. H. Zimerman and A. Villani,
Rev. Bras. Fis. {\bf 7}, 663 (1977).
\bibitem{nami1} B. F. Svaiter and N. F. Svaiter, Phys. Rev. {\bf D47}, 4581 (1993).
\bibitem{nami2} B. F. Svaiter and N. F. Svaiter, J. Math. Phys. {\bf 35}, 1840 (1994).
\bibitem{luis} L. A. Correa-Borbonet, {\em{``Bekenstein
Bound and spectral geometry"}}, ArXiv hep-th/0705.2373 (2007).
\bibitem{bord}  M. Bordag, E. Elizalde, K. Kirsten and S. Leseduarte, Phys. Rev. {\bf D56},
4896 (1997).
\bibitem{vass} M. Bordag, K. Kirsten and D. V. Vassilevich, Phys. Rev. {\bf D59},
085011 (1999).
\bibitem{taka} Y. Takahashi and H. Shimodaire, Nuovo Cimento {\bf
62A}, 255 (1969).
\bibitem{efimov} G. V.
Efimov, {\em{``Strong-Coupling Regime in QFT"}}, Proceedings of
the International Workshop on Quantum Systems, edited by A. O.
Barut, I. D. Feranchuk, Ya. M. Shnir and L. M. Tomil'chik, World
Scientific, Singapure (1994).
\bibitem{proximo} M. Aparicio Alcalde, G. Menezes and N. F.
Svaiter, {\em{``Bekenstein Bound in Quantum Electrodynamics in the
Strong Quantum Fluctuation Regime"}}, in preparation.


\end{thebibliography}
\end{document}